%% file: main.tex
\documentclass[sigconf]{acmart}
\AtBeginDocument{%
  }

\usepackage{multirow,tabularx}
\usepackage{subcaption}
\usepackage{array}
\usepackage{arydshln}
\usepackage{enumitem}
\usepackage[super]{nth}

\copyrightyear{2025}
\acmYear{2025}
\setcopyright{cc}
\setcctype{by-nc-sa}

\acmConference[UMAP '25]{33rd ACM Conference on User Modeling, Adaptation and Personalization}{June 16--19, 2025}{New York City, NY, USA}
\acmBooktitle{33rd ACM Conference on User Modeling, Adaptation and Personalization (UMAP '25), June 16--19, 2025, New York City, NY, USA}
\acmDOI{10.1145/3699682.3728333}
\acmISBN{979-8-4007-1313-2/2025/06}

\begin{document}

\title[Familiarizing with Music: Discovery Patterns for Different Music Discovery Needs]{Familiarizing with Music: Discovery Patterns for Different Music Discovery Needs}


\author{Marta Moscati}
\email{marta.moscati@jku.at}
\orcid{0000-0002-5541-4919}
\authornote{Work carried out during internship at Deezer. Contact Author.}
\affiliation{%
  \institution{Institute of Computational Perception, Johannes Kepler University Linz}
  \streetaddress{Altenberger Straße 69}
  \city{Linz}
  \country{Austria}
}

\author{Darius Afchar}
\orcid{0000-0002-4315-1461}
\email{research@deezer.com}
\affiliation{%
  \institution{Deezer Research}
  \city{Paris}
  \country{France}
}

\author{Markus Schedl}
\email{markus.schedl@jku.at}
\orcid{0000-0003-1706-3406}
\affiliation{
  \institution{Institute of Computational Perception, Johannes Kepler University Linz and Human-centered AI Group, AI Lab, Linz Institute of Technology}
  \city{Linz}
  \country{Austria}
}

\author{Bruno Sguerra}
\orcid{0000-0003-1158-9095}
\email{research@deezer.com}
\affiliation{%
  \institution{Deezer Research}
  \city{Paris}
  \country{France}
}
\newcommand{\embdim}{N}
\newcommand{\eg}{e.\,g., }
\newcommand{\ie}{i.\,e., }
\newcommand{\wrt}{w.\,r.\,t.\, }
\newcommand{\cf}{cf.\, }
\newcommand{\st}{s.\,t.\, }
\newcommand{\cluster}{cluster}
\newcommand{\clusters}{clusters}
\newcommand{\Cluster}{Cluster}
\newcommand{\Clusters}{Clusters}

\newcommand{\question}{``\textit{I'm intrigued by musical styles I'm not familiar with and want to find out more.}''}

\newcommand{\levelofinterest}{\text{IUM}^u}
\newcommand{\trackrank}{\text{ExpR}^u_m}
\newcommand{\relativepop}{\delta P_m}
\newcommand{\relativedist}{\delta D_m}
\newcommand{\diversity}{\text{div}^u}
\newcommand{\stabilityentropy}{\text{SP}_\text{E}^u}
\newcommand{\stabilitynorm}{\text{SP}_\text{N}^u}

\newif\ifworkinprogress
\workinprogressfalse

\ifworkinprogress
	\newcommand{\ms}[1]{\textcolor{blue}{{[Markus] #1}}}
	\newcommand{\mm}[1]{\textcolor{olive}{{[Marta] #1}}}
    \newcommand{\newtext}[1]{\textcolor{purple}{#1}}    
    \newcommand{\deezer}{xxxx}
    \newcommand{\project}{yyyy}
    
\else
    \newcommand{\ms}[1]{}
    \newcommand{\mm}[1]{}
    \newcommand{\newtext}[1]{#1}    
    \newcommand{\deezer}{Deezer}
    \newcommand{\project}{RECORDS,\footnote{RECORDS web page: \url{https://records.huma-num.fr/}} funded by the French National Agency for Research}
\fi

\newcolumntype{H}{>{\setbox0=\hbox\bgroup}c<{\egroup}@{}}

\renewcommand{\shortauthors}{Moscati et al.}

\begin{abstract}
 
Humans have the tendency to discover and explore. This natural tendency is reflected in data from streaming platforms as the amount of previously unknown content accessed by users. Additionally, in domains such as that of music streaming there is evidence that recommending novel content improves users' experience with the platform. Therefore, understanding users' discovery \textit{patterns}, such as the amount to which and the way users access previously unknown content, is a topic of relevance for both the scientific community and the streaming industry, particularly the music one. 
Previous works studied how music consumption differs for users of different traits and looked at diversity, novelty, and consistency over time of users' music preferences. However, very little is known about how users \textit{discover} and  \textit{explore} previously unknown music, and how this behavior differs for users of varying discovery \textit{needs}. 
In this paper we bridge this gap by analyzing data from a survey answered by users of the major music streaming platform \deezer{} 
in combination with their streaming data. We first address questions regarding whether users who declare a higher interest in unfamiliar music listen to more diverse music, have more stable music preferences over time, and explore more music within a same time window, compared to those who declare a lower interest. We then investigate which type of music tracks users choose to listen to when they explore unfamiliar music, 
identifying clear patterns of popularity and genre representativeness that vary for users of different discovery needs. 

Our findings open up possibilities to infer users' interest in unfamiliar music from streaming data as well as possibilities to develop recommender systems that guide users in exploring music in a more natural way
.
\end{abstract}




\maketitle
\section{Introduction} 

    The question of whether recommendations should reinforce users' established preferences or support users in developing their tastes is a long debated one~\cite{resnick2013bursting_filter_bubble,vanchinathan2014explore_exploit}, especially in the domain of music consumption~\cite{celma2016explore_exploit_music}. Research has shown that music listeners tend to like familiar more than unfamiliar music~\cite{ward2014familiarity}, and that users of music streaming platforms tend to listen to music they are familiar with, more often than compared to unfamiliar one~\cite{conrad2019extreme_relistening,tsukuda2020explainable_repeat_consumption}. However, there is scientific evidence that humans have the natural tendency to explore~\cite{mcallister1982variety_seeking} and that music liking initially increases as a function of familiarity, but reaches a peak and decreases for higher levels of familiarity with the music~\cite{reiterhaas2021relistening,sguerra2023ex2vec,sguerra2022discovery_dynamics}. A high-quality music recommendation list should, therefore, find balance between music that is familiar and music that is unfamiliar to the user. Achieving this balance requires a deeper understanding of how music listeners explore new music, which is not only important for the scientific community but also highly relevant from an industry perspective. 
    
    In light of these considerations, previous works~\cite{liang2019exploration_tool,liang2021genre_exploration,liang2023promoting_music_exploration,ferwerda2016personality_traits_diversity,ferwerda2016diversity_cultural,lu2018diversity_adjusting_personality} investigated music exploration behavior by analyzing aspects related to 
    the amount of newly listened content and its diversity 
    defined either at track, artist, or genre level. Often these aspects are analyzed in relation to users' traits. However, these works often lacks granularity regarding several aspects. First, 
    the exploration behavior is not analyzed with respect to the users' self-declared need to discover new music. Although several works considered users' personality traits~\cite{,bonnevilleroussy2013age_music,ferwerda2015personality_music_taxonomy,greenberg2016the_song_is_you,langmeyr2012what_music_personality,ferwerda2016personality_traits_diversity,lu2018diversity_adjusting_personality,rentfrow2003doremi_personality,schaefer2017personality,melchiorre2020personality_correlates}, including openness, these refer to users' general personality and not to music listening specifically. Focusing more on the musical expertise, several works analyzed music listening behavior in relation to users' musical sophistication~\cite{ferwerda2019sophistication_exploration,liang2023promoting_music_exploration}, 
    typically measured through the Goldsmiths Musical Sophistication Index (Gold-MSI)~\cite{muellensiefen2014goldmsi}. This index, however, also captures aspects that go beyond users' self-declared interest in discovering new music, such as their experience as musicians or the money spent on music-related activities. In addition, the analyses often do not consider 
    how the exploration is carried out over time, \ie the different discovery patterns. For instance, they neglect aspects related to which tracks a user listens to while familiarizing with a music genre they previously did not listen to. 
    
    This work bridges these gaps by analyzing the relationship between users' discovery needs and their discovery patterns. We refer to \textit{discovery} as the act of listening to music that was not listened to before, and to \textit{exploration} as the process of discovering tracks of a previously unknown music genre. We consider users' self-reported interest in unfamiliar music as an explicit manifestation of their discovery \textit{needs}. We analyze users' discovery \textit{patterns} in terms of several aspects. First, we look at the amount of newly discovered music, the diversity, and stability of music preferences over time. We then analyze the popularity and genre representativeness of tracks listened \textit{while exploring}. Our aim is to answer the following research questions:   
    \begin{description}[leftmargin=2mm,topsep=0pt]
        \item[RQ1] Do users who declare a higher interest in unfamiliar music \textit{actually do} discover a larger amount of music 
        compared to those with lower declared interest? 
        \item[RQ2] Previous work showed that users with a higher musical sophistication exhibit more diverse and more stable music preferences over time. 
        Does this also apply specifically to users who declare a higher interest in unfamiliar music? 
        \item[RQ3] When users explore 
        music they were not familiar with before, such as a newly discovered genre,  do they start with tracks that are more popular or more representative of the genre, compared to other tracks of the same genre? Does this tendency vary over users declaring different levels of interest in unfamiliar music? 
    \end{description}

    To address these questions, we carry out an extensive analysis on data from the major music streaming platform \deezer{}.\footnote{\url{https://www.deezer.com/en/}} This data consists of records of music listening events from the platform as well as data on users' self-reported interest in listening to unfamiliar music as collected by means of an online survey sent to users via email.

    Our findings show that users that declare a higher interest in unfamiliar music are more active in discovering music
    . They also show that users explore previously unknown music clusters through tracks that are more popular and less representative of genres, compared to other tracks of the explored music. Furthermore, the popularity of listened tracks 
    decreases and the representativeness increases while users keep exploring, and these effects vary for users of different levels of self-reported interest in unfamiliar music. Overall, these results provide valuable insights  into modeling music discovery. The uncovered relationship between self-reported interest in unfamiliar music and discovery patterns are valuable for music streaming platforms that want to guide users in discovering new genres in a more personalized way.

    The rest of the paper is organized as follows: We review previous work related to ours in Section~\ref{sec:background}, discussing separately music exploration and users' music preferences in relation to their traits. We introduce the reader to our terminology and describe our methodology in Section~\ref{sec:methodology}. The experimental setup is described in Section~\ref{sec:experiments}, which also provides an overview of the survey data and of the streaming data used in our analysis. We present the results in Section~\ref{sec:results}, addressing each research question in a separate subsection. Finally, in Section~\ref{sec:conclusions} we contextualize our observations in relation to those of previous work, discuss the limitations and possible extensions of our work.\footnote{The code linked to this work is available at \url{https://github.com/hcai-mms/familiarizing_with_music}, the dataset is linked in the repository.}

\section{Background and Related Work}
    \label{sec:background} 
    \subsection{Music Exploration}
        Early works based on radio listening showed that familiar music tends to be liked more than unfamiliar music~\cite{ward2014familiarity}. More recently, users of music streaming platforms have been shown to stream music they are familiar with more often than unfamiliar one~\cite{conrad2019extreme_relistening,tsukuda2020explainable_repeat_consumption}. These evidences would indicate that recommending music the user is already familiar with is profitable for music streaming platforms and have also motivated the development of music Recommender Systems (RSs) that leverage models of human memory from cognitive architectures~\cite{reiterhaas2021relistening,moscati2023actr,tran2024actr}. However, several works showed that music liking initially increases as a function of ``unfamiliarity'', it reaches a peak and then decreases for higher levels of unfamiliarity~\cite{sguerra2023ex2vec,sguerra2022discovery_dynamics}. Furthermore, recommending novel content has been shown to be a direct source of users' perceived usefulness and satisfaction with recommendations~\cite{castells2022novelty_diversity_rshandbook,pu2011resque}. The purpose of RSs in online streaming platforms therefore does not restrict to reinforcing users' preferences by recommending them familiar content~\cite{resnick2013bursting_filter_bubble,vanchinathan2014explore_exploit,celma2016explore_exploit_music}. This is the case since users have needs to broaden their tastes, a need particularly emerging in the domain of music~\cite{liang2023promoting_music_exploration} which motivated the development of many tools for music discovery~\cite{bostandjiev2012exploration_tool,taramigkou2013exploration_tool,kamalzadeh2016exploration_tool,liang2019exploration_tool,knees2020exploration_tool,schedl2020exploration_tool,cai2021exploration_tool,liang2021interactive_exploration_mood,liang2021genre_exploration,liang2022longitudinal_effects_nudging,petridis2022tastepaths,liang2023promoting_music_exploration,melchiorre2023emomtb}. Additionally, motivated by psychological models of curiosity and exploratory behavior~\cite{berlyne1950exploratory,berlyne1954experimental,berlyne1954theory,berlyne1971aesthetics}, several researchers proposed psychology-informed RSs that define for each user an ideal level of unfamiliarity and provide recommendations that match the ideal unfamiliarity level~\cite{dossantos2015hybrid,zhao2016pcm,zhe2023curiosityrs}. In order to uncover patterns in the exploratory behavior of users of music streaming platforms, several works~\cite{liang2021genre_exploration,liang2023promoting_music_exploration} analyzed the consistency of music preferences in terms of repeated consumption at either artist, genre, or track level, and at different time scales.  
        
        
        There is, however, a lack of insight on \textit{how} users explore a music genre, after they are exposed to it for the first time. 
        In our work we therefore take a different perspective: We aim at analyzing which tracks users listen to when exploring 
        newly discovered music
        . We analyze this under different perspectives, such as popularity and genre representativeness of the music tracks, and as a function of the distance in time from the first exposure to similar music. 
        
    \subsection{Music Preference and User Traits}
        The question of which individuals' traits are associated with specific music preferences or listening behaviors is a long-debated one. Previous studies explored factors related to age~\cite{bonnevilleroussy2013age_music,ferwerda2017personality_age,greenberg2016the_song_is_you}, gender~\cite{greenberg2016the_song_is_you,mccown1997music_personality_gender}, cognitive style~\cite{greenberg2015music_cognitive_style}, cultural~\cite{ferwerda2016soap,ferwerda2016diversity_cultural,skowron2017genre_preferences_culture_socioeconomic}, and personality traits~\cite{hu2010personality_recommendation_perception,bonnevilleroussy2013age_music,ferwerda2015personality_music_taxonomy,greenberg2016the_song_is_you,langmeyr2012what_music_personality,ferwerda2016personality_traits_diversity,lu2018diversity_adjusting_personality,rentfrow2003doremi_personality,schaefer2017personality,melchiorre2020personality_correlates}. Taking an approach based solely on the streaming data, Villermet et al.~\cite{villermet2021follow_the_guides} also showed that the diversity of music consumption is affected by users' behavioral patterns in the way they choose to access music (\eg through algorithmic recommendations, editorial playlists, personal playlists, \dots). Closer to our work, several works focused on listeners' musical sophistication measured with the Gold-MSI~\cite{muellensiefen2014goldmsi}. For instance, Ferwerda and Tkalčič~\cite{ferwerda2019sophistication_exploration} showed that users' musical sophistication correlates with the diversity in music consumption. To further understand music preferences and their development for users of different musical sophistication, Liang and Willemsen carried out several studies on the listening behavior of Spotify users~\cite{liang2019exploration_tool,liang2021genre_exploration,liang2022longitudinal_effects_nudging,liang2023promoting_music_exploration}. More specifically, the authors analyzed users' musical sophistication in relation to their choice of {how far the explored genre should be from their favorite genre}, and to the desired level of personalization in the genre exploration. The authors showed that the selection behavior differs for users with different levels of active engagement according to the the Gold-MSI, pointing out how users with a higher active engagement tend to explore genres that are closer to their favorite one. 
        These results indicate that different levels of musical sophistication 
        lead to different choices on which music to explore. However, the criterion for the choice of genre (\ie the distance from the user's favorite genre) is not a strong indication that the user was not exposed to the genre before, or that they were not familiar with it; for instance, we could expect that users are already familiar with genres that are close to their favorite one. 
        Additionally, Liang and Willemsen's works provide insight into {how} users explore genres focusing on the desired level of personalization. Finally, the active engagement of the Gold-MSI is considered as an aggregated score, combining several aspects, such as the time and money invested in music-related activities. 
        
        Our work is complementary to the ones described above. In particular, we analyze users' discoveries for music \textit{they did not stream before} and therefore are less familiar with. Furthermore, we analyze their discovery \textit{patterns}, \ie how they explore music
        , by also looking at the popularity and genre representativeness of tracks they choose to listen to while exploring new music. Finally, we focus on a single aspect of the active engagement, which we believe to be most related to their  discovery \textit{needs}: the self-reported interest in unfamiliar music as measured by the Gold-MSI.

    \section{Methodology}
        \label{sec:methodology}   
        In this section we describe the terminology and methodology used in our work. The terminology is also summarized in Table~\ref{tab:methodology_terminology}.


        \subsection{Users' Interest in Unfamiliar Music}
            We denote users with $u$. Users replied to the question from the Gold-MSI questionnaire \question{} on a $1$ (fully disagree) to $5$ (fully agree) Likert scale as part of a survey sent via email.\footnote{The original Gold-MSI questionnaire is based on a 7-point Likert scale. The scale was converted to a 5-point one to simplify the form of the questionnaire.}  We refer to user $u$'s reply as self-reported Interest in Unfamiliar Music ($\levelofinterest$). We use $\levelofinterest$ as a measure of $u$'s discovery needs.          
        \subsection{Tracks' \Cluster{} Labels}
            \label{sec:methodology_editorial_playlists}
            We denote music tracks with $m$. We identify similar tracks based on their co-occurrence in \textit{user}-generated playlists and assign similar tracks the same label $c$. We use \textit{editorial} playlists to validate the meaning of labels $c$ in terms of music genres and \textit{not} for the definition of the labels.\footnote{Other works rely on genre labels extracted from user-generated tags. We excluded the use of user-generated tags to infer music genres since these are often noisy and prone to vandalism, fraud, or strategies intended to misclassify artists' genres.} 
            To define labels, we apply $k$-means clustering to vector representations of the user-generated playlist data of music tracks. We use as representations for clustering the embeddings 
            computed by decomposing the mutual-information matrix representing song co-occurrences in \deezer{} user-generated playlists, decomposed with singular value decomposition (SVD). We use $\mathscr{l}^2$ as norm for clustering. 
            
            We validate the \cluster{} definitions with information from editorial playlists\footnote{\deezer{} editorial playlists are created by experts of one or more music genres employed at \deezer{} also for the task of identifying tracks that they believe are most representative of the genre they are expert in. To our knowledge, their choices are not influenced or biased by economic incentives related to the promotion of specific content.} created by music experts at \deezer{}.\footnote{Notice that the embeddings used for clustering are based on a larger set of user-generated playlists, excluding playlists generated with RSs. The smaller set consisting of editorial playlists is only used to investigate the meaning of the \cluster{} labels.} More specifically, we want to check that different cluster labels $c$ reflect different music genres, and that tracks closer to cluster centroids are more representative of music genres.  
            For the genres \{Pop, Rock, Classical, Electronic\},\footnote{We selected Pop, Rock, and Electronic since these are the three genres with the highest number of recordings according to data provided by Discogs (\url{https://mtg.github.io/acousticbrainz-genre-dataset/data_stats/}). We chose to extend this with classical music since it provides a genre that is often complementary to more commercial ones.} we select the editorial playlists named ``\textit{<genre> Essentials}''. We label each editorial playlist with the cluster labels $c$'s of its tracks, and analyze the similarity between editorial playlists in terms of Jaccard index computed over cluster labels. {Labels that are not representative of music genres result in clusters evenly spread across editorial playlists and high Jaccard similarity. In contrast, labels that accurately represent genres lead to lower similarity between playlists.}  We select the number of \clusters{} $n_\text{clusters}$ that ensures cluster homogeneity, measured both as the Within-Cluster Sum of Squares (WCSS), and as low Jaccard similarity of the editorial playlists.\footnote{We do not control for cluster size since clusters are reflective of music genres and since music genres come with different numbers of tracks
            .}  Furthermore, we label with $D_m$ the distance of track $m$ from the centroid of the cluster it belongs to. For each cluster $c$, we then compute the average distance $\bar{D}_c$ over tracks in $c$, including tracks that are and tracks that are \text{not} in the editorial playlists. $\bar{D}_c$  represents the distance from cluster centroid ``expected`` for tracks of cluster $c$. For each track in $c$, we compute the relative deviation in distance from the cluster centroid as $\relativedist = (D_m - \bar{D}_c) / \bar{D}_c$. We then average $\relativedist$ for tracks $m$ in the editorial playlists. 
            By doing so, an average over tracks in the editorial playlists of $\relativedist = -100\%$ indicates that these tracks are always located at the centroid, while an average of $\relativedist = 0\%$ indicates that the average over tracks in editorial playlists is equal to the average over all tracks in the cluster.
            
            We find\footnote{We refer the reader to Section~\ref{sec:experiments} for more details.} that $\relativedist < 0$, which indicates that tracks in editorial playlists are closer to the cluster centroids compared to tracks that are not in these playlists. Since editorial playlists are curated to reflect a given music genre for multiple users, this finding allows us to consider the deviation in the relative distance from cluster centroids $\relativedist$ as a measure of the ``representativeness'' of track $m$ of the music genre of the \cluster{} $c$ it belongs to.


        \subsection{Newly Discovered Music \Clusters{}}
            \label{sec:methodology_previously_unexplored_music}

            We say that user $u$ \textit{discovered} track $m$ if they listened to it for the first time, and that user $u$ \textit{discovered} \cluster{} $c$ if they listened for the first time to a track labeled with $c$. We refer to user's $u$ \textit{exploration} of \cluster{} $c$ as the time-ordered list of first tracks that $u$ listens from \cluster{} $c$. To identify newly discovered music tracks and \clusters{}, we first define the long-term preferences of each user as the tracks and \clusters{} they already listened from the timestamp $t_0$ of their first interaction to a timestamp $t_1$ corresponding to a time-window $[t_0, t_1]$ of six months of listening. We select a six-month window as this is also used by Mehrotra et al.~\cite{mehrotra2021spotify_familiarity} to define discoveries as tracks that a user has not listened before.  For each user $u$, we measure the ratio of known tracks $n^u_c$ that belong to \cluster{} $c$, $p^u_{c} = n^u_c / \sum_c n^u_c$; this ratio varies with time, as users discover new tracks. We therefore label this with $p^u_c(t)$ and define $u$'s long-term preferences $\mathbf{p}^u(t_1)\in \mathbb{R}^{n_\text{clusters}}$ as the vector of ratios $p^u_c(t_1)$ of known tracks per \cluster{} $c$ at the end of the six-months window $[t_0, t_1]$. For the remaining listening events, \ie those happening at $t> t_1$, we define the time of discovery of a new \cluster{} $c$ as the timestamp of first listen of a track from \cluster{} $c$, \ie the timestamp $t^*$ such that $p^u_c(t) = 0$ for $t<t^*$ and $p^u_c(t) > 0$ for $t\geq t^*$. We count the number $N^u$ of distinct \textit{new} \clusters{} discovered in the one-week window $[t_1, t_2]$ following the initial six-months. To measure how the amount of discovered music varies with $\levelofinterest$, we first  
            compute Pearson's correlation coefficient\footnote{Using Spearman’s rank correlation leads to similar results (see repository).} between $N^u$ and $\levelofinterest$ over all users. Furthermore, we apply bootstrapping to estimate the mean and standard error of $N^u$ for groups of users of different $\levelofinterest$ and plot these as a function of $\levelofinterest$. We take 1{,}000 samples of sample size 100, sampling with replacement. 
            
        \subsection{Diversity and Stability of Music Preferences}
            \label{sec:methodology_diversity_stability}
            We quantify the diversity and time evolution of music preferences in terms of relative amounts of known tracks per \clusters{}, $p^u_c(t)$. We measure diversity~\cite{ferwerda2016soap,ferwerda2019sophistication_exploration,liang2023promoting_music_exploration} as Shannon's entropy\footnote{We refer the reader to the repository for an analysis in terms of Rao-Stirling diversity~\cite{way2019diversity}, leading to analogous results. We expect similar results also when using GS-score as diversity measure, due to the high correlation of the two metrics~\cite{anderson2020diversity_spotify}.} over \clusters{}, $\diversity(t) = -\sum_c p^u_c(t) \log p^u_c(t)$. We measure the time evolution of music preferences as time derivative\footnote{We approximate the derivative with the difference quotient.} of $\diversity(t)$ and of $\left\lVert \mathbf{p}^u(t)\right\rVert_2$, \ie the $\mathscr{l}^2$ norm of the vector $\mathbf{p}^u(t)\in \mathbb{R}^{n_\text{clusters}}$ representing the relative amount of known tracks per \clusters{}. We name the resulting quantities Stability Preference with Entropy ($\stabilityentropy$) and Stability Preference with Norm  ($\stabilitynorm$), respectively. 
            
            Both $\stabilityentropy$ and $\stabilitynorm$ measure how $u$'s diversity in music preferences evolves in time. Positive values of $\stabilityentropy$ indicate that $u$'s diversity is increasing (increase in $\diversity$ over time), while positive values of $\stabilitynorm$ that $u$'s diversity is decreasing (towards $\mathbf{p}^u(t)$ being directed along one $p^u_c(t)$ axis only).\footnote{Notice that $\left\lVert \mathbf{p}^u(t)\right\rVert_2^2$ corresponds to Simpson's index
            , whose inverse is also often used as diversity measure~\cite{ferwerda2019sophistication_exploration}.}
            
            The six-months window of $u$'s initial listening events $[t_0, t_1]$ is first used to determine the baseline of $u$'s preferences, $\mathbf{p}^u(t_1)\in \mathbb{R}^{n_\text{clusters}}$, \ie the vector of ratios $p^u_c(t_1)$ of known tracks per \cluster{} $c$ at time $t_1$. We then average $\stabilityentropy$ and $\stabilitynorm$ over the listening events of $u$ for $t \geq t_1$ to keep into account artifacts of the beginning of the streaming data
            . 
            
            To measure how the diversity and stability of music preferences vary with $\levelofinterest$, we compute Pearson's correlation coefficient\footnote{Using Spearman’s rank correlation leads to similar results (see repository).} between $\diversity(t_1)$ (the diversity after six months of listening), $\stabilityentropy$, and $\stabilitynorm$ with $\levelofinterest$ over all users. We then apply bootstrapping to estimate the mean and standard error of $\diversity(t_1)$, of $\stabilityentropy$, and of $\stabilitynorm$ within each group of users $\levelofinterest\in [1, \dots, 5]$ and plot these as a function of $\levelofinterest$. We take 1{,}000 samples of sample size 100, sampling with replacement. 
            
        \subsection{Time Exploration of Music \Clusters{}}
            \label{sec:methodology_exploration_behavior}
            We first explore how popularity and \cluster{} representativeness 
            evolve over the first tracks listened when a music \cluster{} is explored, 
            and over all users, \ie irrespective of $\levelofinterest$. Building upon the analysis on the editorial playlists described in Section~\ref{sec:methodology_editorial_playlists}, we measure the \textit{\cluster{} representativeness} of track $m$ in $c$ as $\relativedist$, \ie the deviation in its $\mathscr{l}^2$ distance from the centroid of cluster $c$ relative to the average distance of tracks in $c$. We define the \textit{popularity} $P_m$ of track $m$ 
            as the logarithm of the total number of its listening events; in a similar way to $\relativedist$, we define $\relativepop = (P_m - \bar{P}_c) / \bar{P}_c$ as the relative deviation of $m$'s popularity from the average popularity of tracks of the same cluster $c$. We do so to keep into account that different clusters might have different distributions of popularity and distance from cluster centroid. 

            When user $u$ discovers cluster $c$, we assign each track $m$ a value according to its Rank in the EXPloration ($\trackrank$) of $c$. By doing so, $\trackrank=1$ for the first track $u$ listened from $c$, $\trackrank=2$ for the second track, and so on.\footnote{We stop at rank $\trackrank = 10$, for which the analyzed trends stabilize and refer the reader to the repository for higher ranks. } To measure how $\relativepop$ and $\relativedist$ vary with $\trackrank$ we compute Pearson's correlation between $\trackrank$ and both $\relativepop$ and $\relativedist$. To display the evolution of $\relativedist$ graphically, we apply principal component analysis (PCA) to the SVD embeddings of all tracks. Then, for each value of $\trackrank$, the distribution of distances from the cluster centroid of the first two principal components is approximated with kernel density estimation (KDE) using a Gaussian kernel. The KDE is used to plot the isolines of the density for the values 0.2, 0.4, and 0.6\%. These isolines therefore give an overview of the genre representativeness in the exploration of \clusters{}, and across all users. For each value of $\trackrank$, we apply bootstrapping to estimate the mean and standard error of the distribution of $\relativepop$ of tracks of rank $\trackrank$. We take 1{,}000 samples of sample size 100, sampling with replacement. Finally, to analyze the exploration of music \clusters{} for different levels of $\levelofinterest$, we perform two multivariate linear regressions, one of $\relativepop$ and one of $\relativedist$, as a function of both the time-rank $\trackrank$ of the explored track, and of $\levelofinterest$.


            \begin{table}
                    \scalebox{0.86}{
                \begin{tabular}{rp{.38\textwidth}}
                    \toprule
                    Variable & Description \\
                    \midrule
                    $\levelofinterest$ & $u$'s  interest in unfamiliar music \ie reply to ``\textit{I'm intrigued by [...] more.}'' on a $1$ (fully disagree) to $5$ (fully agree) Likert scale. \\
                    $N^u$ & Number of distinct \clusters{} discovered by $u$ in a one-week window.\\
                    $\relativepop$ & Popularity of $m$ relative to the average popularity of its cluster $c$. $\relativepop=0\%$ means that $m$ has popularity equal to $c$'s average, $\relativepop>0\%$ that $m$ is more popular than $c$'s average.\\
                    $\relativedist$ & Distance of $m$ from the centroid of its cluster $c$ and relative to the average distance of tracks in $c$. $\relativedist = 0\%$ if $m$'s distance from the centroid is equal to the average in $c$, $\relativedist > 0\%$ if $m$ is further away from the centroid and less representative of $c$. \\
                    $p^u_c(t)$ &  Ratio of tracks known to $u$ that are in $c$ at time $t$.\\
                    $\mathbf{p}^u(t)$ & $u$'s preferences at time $t$ expressed as vector of ratios of known tracks $p^u_c(t)$. \\
                    $\diversity(t)$ & Diversity of $u$'s music preferences at time $t$ measured as Shannon's entropy over components of $\mathbf{p}^u(t)$.\\
                    $\stabilityentropy$ & Stability of $u$'s music preferences measured as time derivative of $\diversity(t)$, averaged over time. \\
                    $\stabilitynorm$ &  Stability of $u$'s music preferences measured as time derivative of $\left\lVert \mathbf{p}^u(t)\right\rVert_2$, averaged over time. \\
                    $\trackrank$ & $m$'s rank in $u$'s exploration of 
                    cluster $c$. $\trackrank=1$ for the first track listened by $u$ in $c$.\\
                    \bottomrule
                \end{tabular}}
                \caption{Summary of the terminology. $u$ indicates a user, $m$ a music track, $t$ a timestamp.}
                \label{tab:methodology_terminology}
            \end{table}

    \section{Experimental Setup}
        \label{sec:experiments}
\begin{figure}
                \centering
                \includegraphics[width=0.48\textwidth]{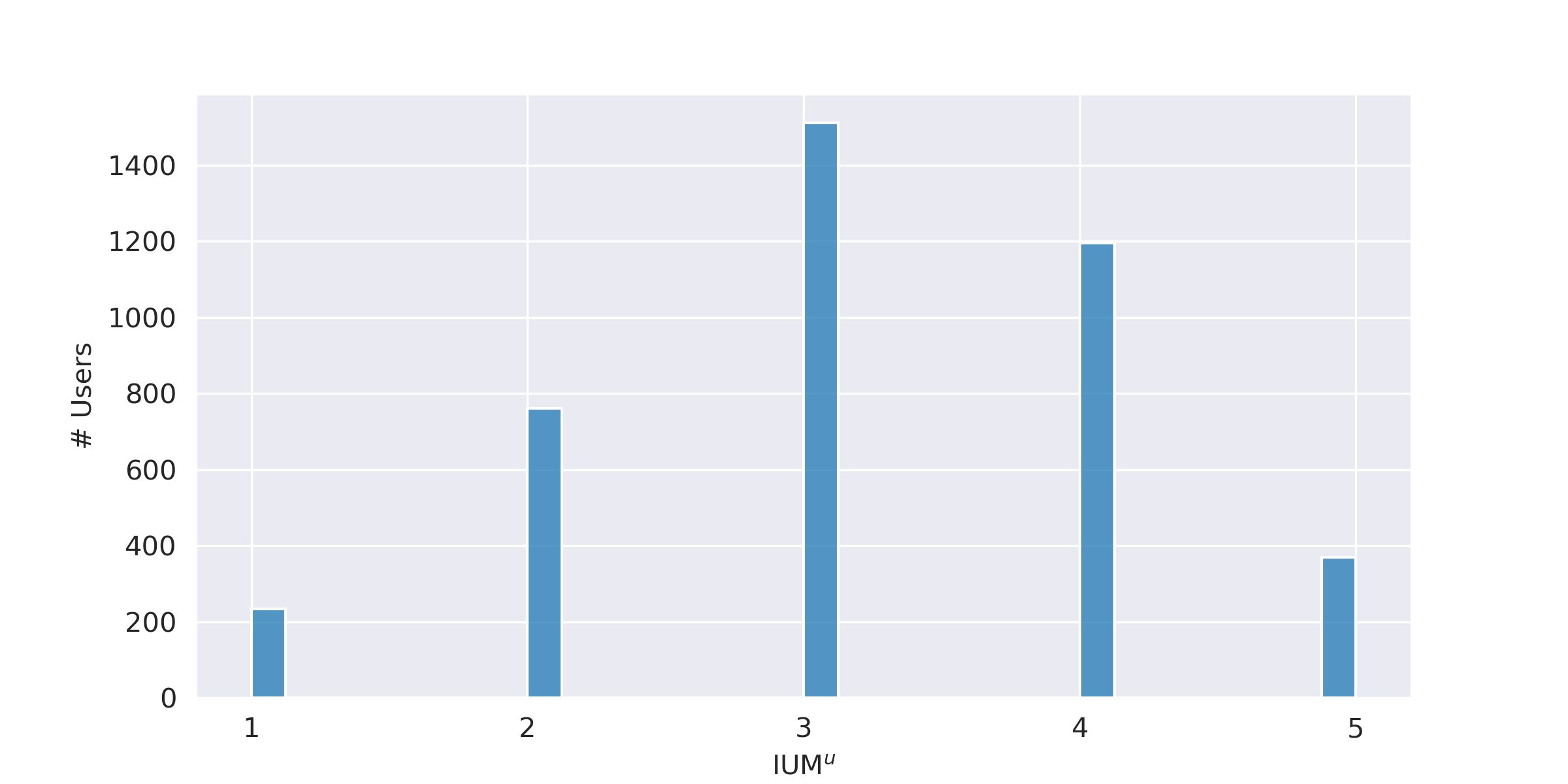}
                \caption{Histogram of $\levelofinterest$, \ie of replies to the question \question{}. $\levelofinterest=1$: fully disagree; $\levelofinterest = 5$: fully agree.}
                \Description[Histogram of the replies to the question \question{}.]{Histogram of the replies to the question \question{}, \ie self-reported interest in unfamiliar music. $\levelofinterest=1$: fully disagree; $\levelofinterest = 5$: fully agree.}
                \label{fig:survey_replies}
            \end{figure}

            We perform our analyses on a set of users of the music streaming platform \deezer{}. The dataset was collected within the project \project{}, whose aim is to study practices on music streaming platforms. In particular, we consider users that took part in a survey, sent to users via email, regarding their music listening practices, musical tastes and socio-demographic variables. The survey was developed to evaluate users' musical sophistication based on the Gold-MSI questionnaire~\cite{muellensiefen2014goldmsi},\footnote{Gold-MSI web page: \url{https://shiny.gold-msi.org/gmsi_toplevel/}} to which users replied on a $5$-point Likert scale. 
            We assign $u$ a value of $\levelofinterest$ equal to their reply to \question{}. 
            In addition, the dataset provides the listening events of these users from 01/01/2022 till 31/12/2022. We only consider listening events that lasted longer than 30 seconds, a threshold often applied in  music streaming for remuneration purposes, and in music recommendation to identify positive feedback~\cite{jiang2024spotify_threshold,tran2024actr}; to ensure that user $u$ ``familiarized'' with track $m$, we disregard user--track pairs $(u, m)$ for which $u$ listened to $m$ only once in this period. For each user--track pair $\levelofinterest$ we keep the timestamp $t$ of first interaction as indication of when $u$ discovered $m$ (and the respective \cluster{} $c$, if no track from $c$ was listened before). We restrict to users that replied to all questions from the Gold-MSI and that provided demographic information.  We apply 10-core filtering and only restrict to users that listened to at least 10 distinct tracks, and tracks that were listened by at least 10 distinct users. This leads to 4{,}070 users, 72{,}555 tracks, and to 28{,}678{,}935 user--track interactions, corresponding to 3{,}425{,}564 unique user--track pairs  $(u, m)$, as summarized in Table~\ref{tab:interaction_data}. The users constitute of $2{,}207$ male, $1{,}863$ female; the collected replies $\levelofinterest$ had an average of $3.2$ and a standard deviation of $1.0$. Figure~\ref{fig:survey_replies} shows the histogram of $\levelofinterest$ while Table~\ref{tab:user_data} reports, for each gender, the number of users who responded with each allowed value of $\levelofinterest$.           

            We vary the number $n_\text{clusters}$ of \clusters{}  in $\{8, 16, 32, 64, 128, 256\}$. To ensure \cluster{} homogeneity, we select $n_\text{clusters}$ as the smallest value above the elbow in the WCSS plot, and for which the Jaccard index computed over cluster labels between any pair of editorial playlists is below $1/3$, indicating that for any pair of editorial playlists, the overlapping cluster labels are less than $1/3$ of the labels appearing in any of the two playlists. This leads to $n_\text{clusters} = 128$.

           \begin{center}
            \begin{table*}[!htb]
                \begin{minipage}{\columnwidth}
                    \centering
                    
                    \scalebox{0.86}{
                    \begin{tabular}{c|r|r|r|r|r}
                    & $\levelofinterest=1$  & 2 & 3 & 4 & $5$ \\\hline
                    m& 135 ($6\%$) & 393 ($18\%$) & 794 ($36\%$) & 683 ($31\%$) & 202 ($9\%$)\\
                    f& 98 ($5\%$)& 368 ($20\%$)& 717 ($39\%$)& 512 ($27\%$)& 168 ($9\%$)\\
                    \end{tabular}}
                    \caption{Number of users by gender (m, f), and by $\levelofinterest$, \ie reply to the question \question{}. $\levelofinterest=1$: fully disagree; $\levelofinterest = 5$: fully agree. Percentages are expressed relative to participants of the same gender.}
                    \label{tab:user_data}
                \end{minipage}\hfill 
                \begin{minipage}{\columnwidth}
                    \centering
                    \scalebox{0.86}{
                    \begin{tabular}{r|r|r| r}
                    \# Users $u$ & \# Tracks $m$ & \# Interactions & \# Distinct $(u, m)$ pairs \\ \hline 4{,}070 & 72{,}555 & 28{,}678{,}935 &  3{,}425{,}564 \\
                    \end{tabular}}
                    \caption{Characteristics of the streaming data. "\# Interactions" consider listening events individually, counting relistening of a same track as separate events. "\# Distinct $(u, m)$ pairs" aggregates listening events of a same track from a same user, considering relistening as a single pair.}
                    \label{tab:interaction_data}
                \end{minipage}
            \end{table*}
            \end{center}  
     
\section{Results}
    \label{sec:results} 
    Before moving to the results on RQ1, 2, and 3, we provide more insight on the validation of \cluster{} labels $c$ and on $\relativedist$ in relation to \cluster{} representativeness.  
    
    The column ``\textit{\# Editorial Tracks}'' in Table~\ref{tab:cluster_editorial_playlists} shows the number of distinct tracks that are present in the selected editorial playlists,\footnote{This table refers to the validation of \cluster{} labels $c$ and representativeness of centroids. It \emph{does not} describe the full set of tracks of the streaming data used in our analyses. For a summary of the streaming data, we refer the reader to Table~\ref{tab:interaction_data}.} while the column ``\textit{\# Editorial Tracks in the Streaming Data}'' shows the number of tracks appearing both in the selected editorial playlists and in the streaming data used in our analysis. The last column shows the average of $\relativedist$ over tracks appearing both in the selected editorial playlists and in the streaming data used in our analysis. We see that on average tracks that appear in editorial playlists are closer to cluster centroids than other tracks, since for editorial tracks $\relativedist< 0$. This motivates our use of $\relativedist$ as measure of track representativeness of the genre of a \cluster{}. 
    
                \begin{center}
                \begin{table*}[!htb]
                    \scalebox{0.86}{
                    \begin{tabular}{l|r|r|r}
                         Editorial Playlist&\# Editorial Tracks &\# Editorial Tracks in the Streaming Data & Average $\relativedist$ 
                         \\\hline
                         Classical Essentials &23 &  23 & -31\% \\
                         Electronic Essentials & 27 & 18& -7\%\\
                         Pop Essentials & 81 & 55& -15\%\\
                         Rock Essentials & 70 & 65& -21\%\\
                    \end{tabular}}
                    \caption{Number of tracks of the editorial playlists for Classical, Electronic, Pop, and Rock, and of tracks of the streaming data that also appear in the editorial playlists. Last column: $\relativedist$, \ie distance from cluster centroid relative to the mean distance of tracks of the same \cluster{}, averaged over tracks of the editorial playlist.}
                    \label{tab:cluster_editorial_playlists}
                \end{table*}
                \end{center}
    

        \subsection{RQ 1: Newly Discovered Music \Clusters{}}   
            \label{sec:rq1}
\begin{figure}
                \centering
                \includegraphics[width=0.48\textwidth]{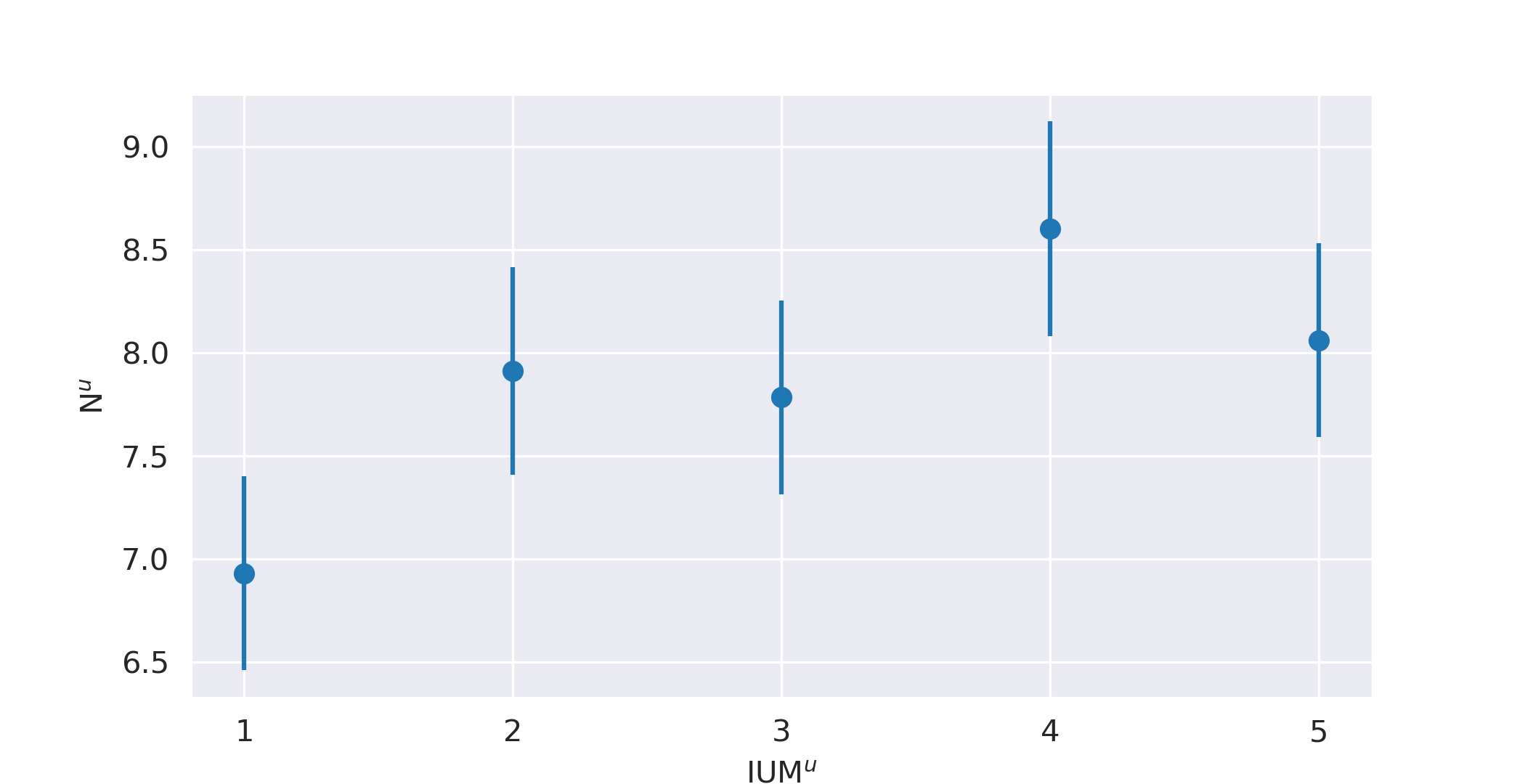}
                \caption{Average number $N^u$ of \clusters{} explored in $[t_1, t2]$ (one week after initial six months) for different values of $\levelofinterest$. Central dots and error bars represent mean and standard errors estimated with bootstrapping (see Section~\ref{sec:methodology_previously_unexplored_music}).}
                \Description[Number of \clusters{} explored in one week for different values of self-reported interest in unfamiliar music $\levelofinterest$.]{Number of \clusters{} explored in one week for different values of self-reported interest in unfamiliar music $\levelofinterest$. Central dots and error bars represent mean and standard errors estimated with bootstrapping (see Section~\ref{sec:methodology_previously_unexplored_music}).}
                \label{fig:rq1_number_explored_genres}
            \end{figure}

            Figure~\ref{fig:rq1_number_explored_genres} shows the number $N^u$ of explored music \clusters{} in $[t_1, t_2]$, \ie within a one-week period after the initial six months $[t_0, t_1]$. The central dots and error bars represent the mean and standard error  of $N^u$ for different values of $\levelofinterest$, computed as described in Section~\ref{sec:methodology_previously_unexplored_music}.
            Overall the plot shows that users of higher $\levelofinterest$ tend to explore, within a same time-window, more music \clusters{} than users of lower $\levelofinterest$. The most remarkable difference is between users of $\levelofinterest=1$ and $\levelofinterest=4$, the latter exploring on average $1.5$ more music \clusters{} than the former, in one week ($N^u\simeq 8.5$ vs. $N^u\simeq 7$). Overall, this trend is confirmed by the positive, albeit low, correlation coefficient ($\rho=0.06$, $p=0.00161$) computed over all users between $N^u$ and $\levelofinterest$, indicating a weak correlation.

        \subsection{RQ 2: Diversity and Stability of Music Preferences}
            \label{sec:rq2}
\begin{figure}
                \centering
                \begin{subfigure}[t]{0.48\textwidth}
                    \centering
                    \includegraphics[width=\textwidth]{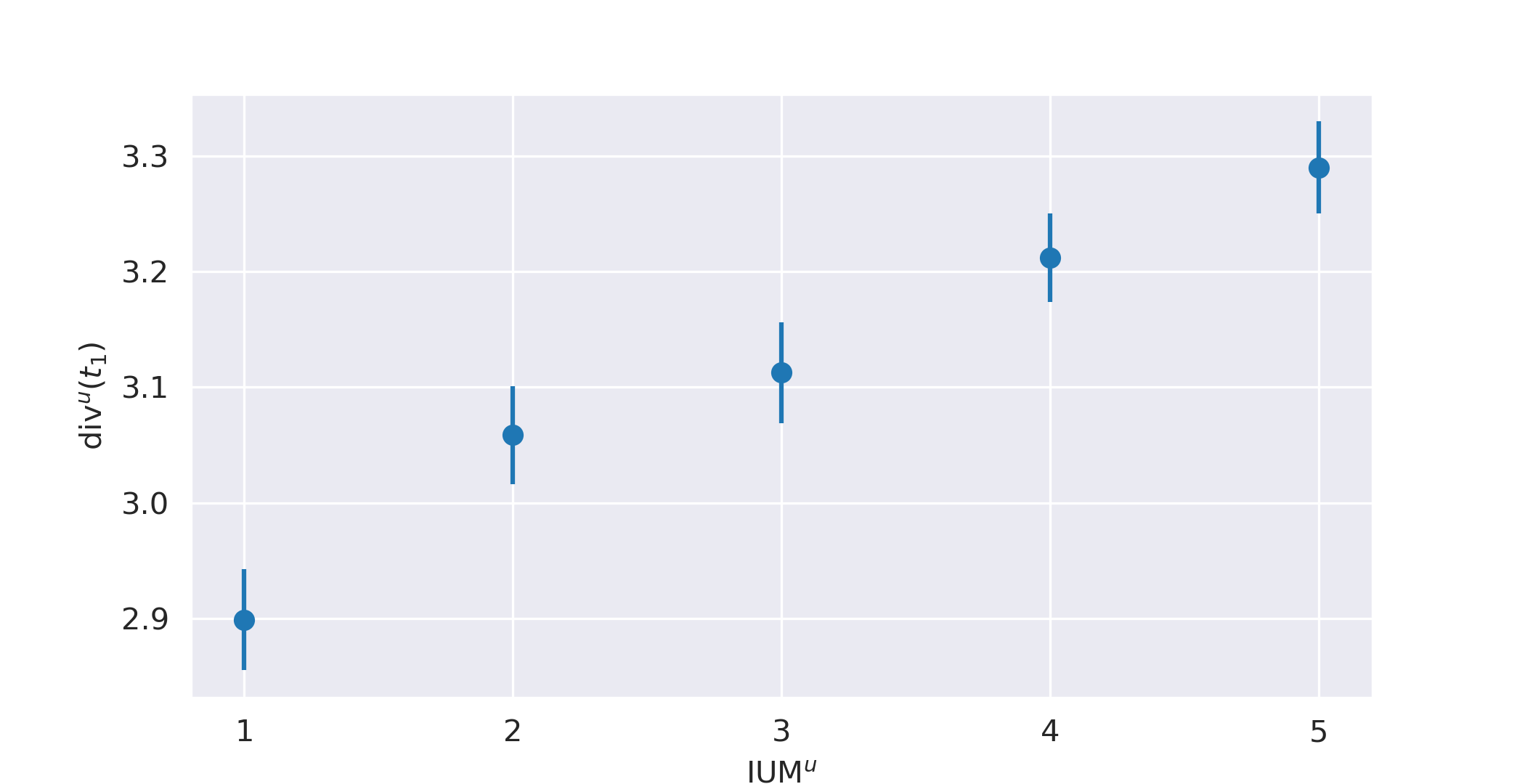}
                    \caption{Diversity $\diversity(t_1)$, \ie Shannon entropy over music \clusters{} after six months listening, for different users' $\levelofinterest$.}
                    \label{fig:rq2_entropy_six_months}
                \end{subfigure}
                \hfill
                \begin{subfigure}[t]{0.48\textwidth}
                    \centering \includegraphics[width=\textwidth]{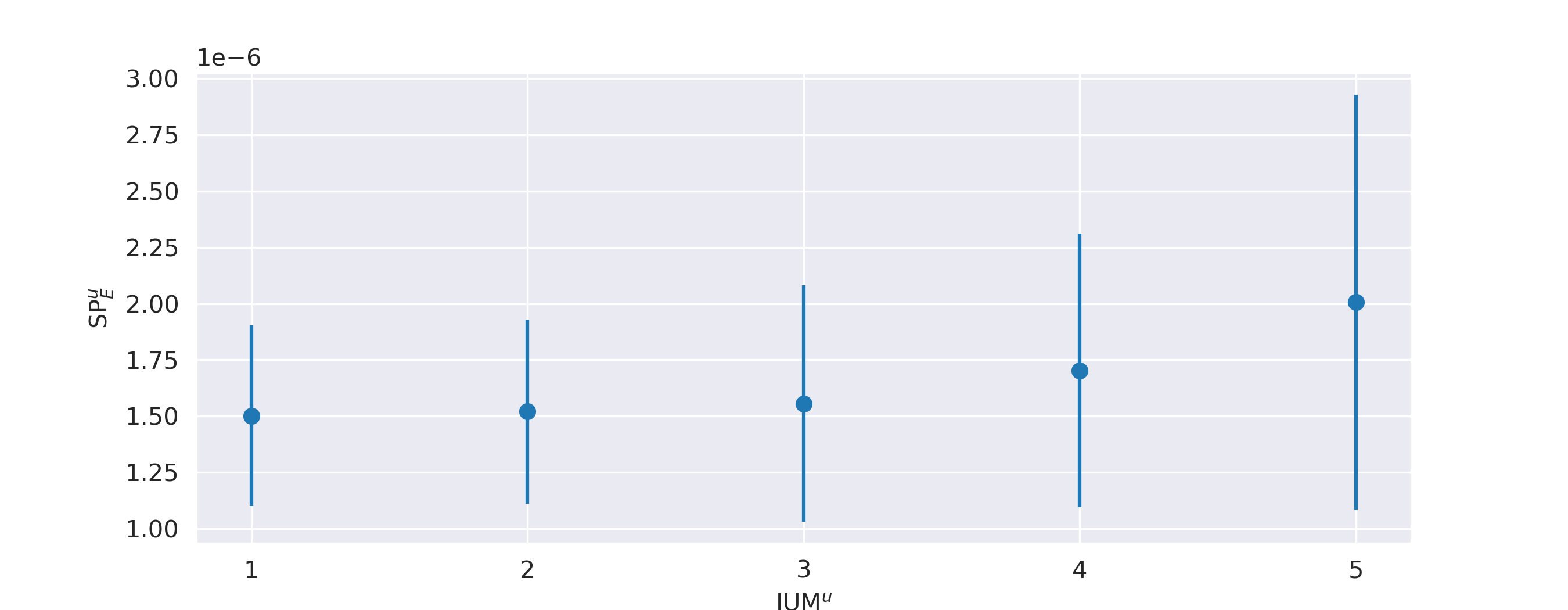}
                    \caption{Music preference stability as $\stabilityentropy$, \ie time-derivative of $\diversity(t)$, for different users' $\levelofinterest$.}
                    \label{fig:rq2_entropy_derivative}
                \end{subfigure}
                \hfill
                \begin{subfigure}[t]{0.48\textwidth}
                    \centering \includegraphics[width=\textwidth]{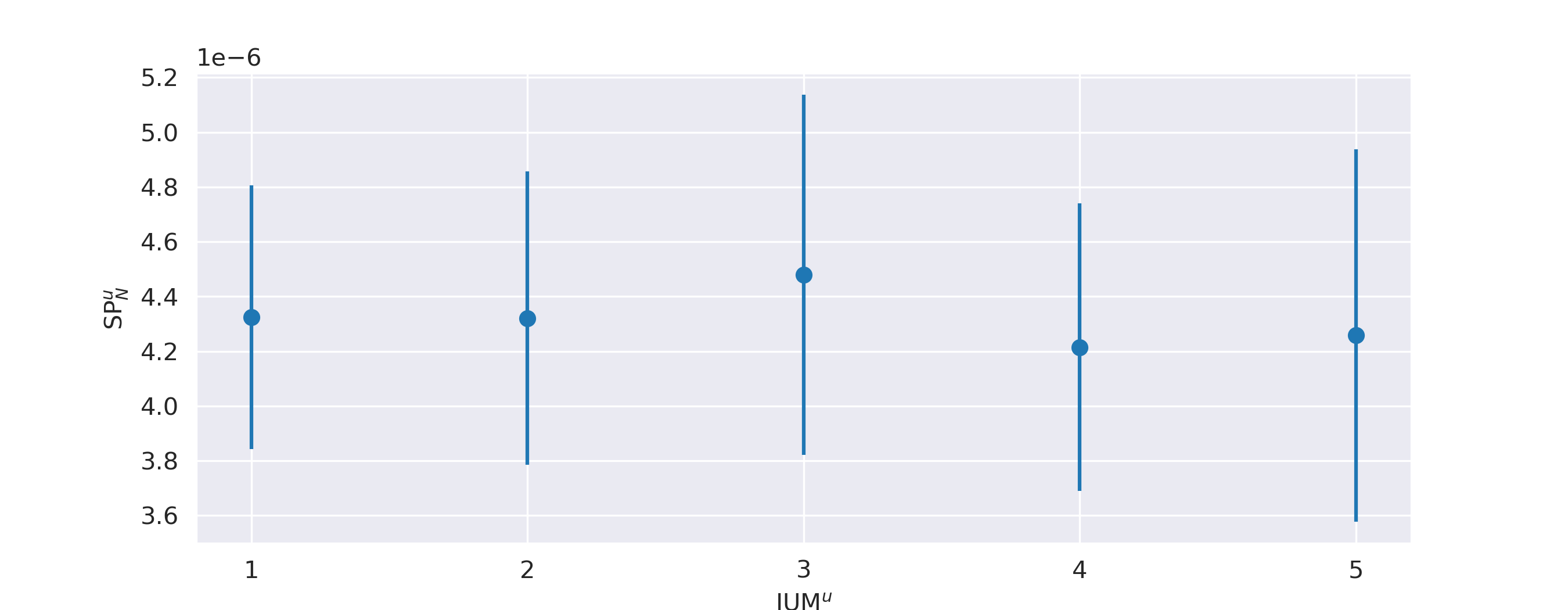}
                    \caption{Music preference stability as $\stabilitynorm$, \ie time-derivative of $\left\lVert \mathbf{p}^u(t)\right\rVert_2$, for different users' $\levelofinterest$.}
                    \label{fig:rq2_probability_derivative}
                \end{subfigure}
                \caption{Diversity and stability of music preferences for different declared interest in unfamiliar music $\levelofinterest$. Central dots and error bars represent the mean and standard errors estimated with bootstrapping.}
                \label{fig:rq2_diversity_entropy_probability_derivative}
                \Description[Diversity and stability of music preferences for users of different declared interest in unfamiliar music $\levelofinterest$.]{Entropy over music genres and Probability Derivative for different declared interest in unfamiliar music.}
            \end{figure}

            The diversity $\diversity(t_1)$ of music consumption computed as entropy over listened \clusters{} at the end of the initial six-months period, is displayed in Figure~\ref{fig:rq2_entropy_six_months}. Higher values of $\diversity(t_1)$ indicate music preferences more equally distributed over different clusters, and lower values indicate music preferences more peaked over few clusters. The figure clearly shows an increasing trend, indicating that users who declare a higher interest in unfamiliar music $\levelofinterest$ also have higher $\diversity(t_1)$, and hence more diverse music tastes. This dependency is confirmed by the positive correlation coefficient computed over all users ($\rho=0.19$, $p<10^{-5}$) between $\levelofinterest$ and $\diversity(t_1)$. It is in line with the results from previous works~\cite{ferwerda2019sophistication_exploration,liang2023promoting_music_exploration}, which grouped users according to their active engagement values of the Gold-MSI, a value to which the self-reported interest in unfamiliar music $\levelofinterest$ also contributes. Notice that this insight differs from the one described in Section~\ref{sec:rq1}: While Figure~\ref{fig:rq1_number_explored_genres} indicates that users with higher interest in unfamiliar music explore \textit{more} \clusters{}, Figure~\ref{fig:rq2_entropy_six_months} indicates that they have \textit{more evenly spread} interests over \clusters{}.

            Figures~\ref{fig:rq2_entropy_derivative} and~\ref{fig:rq2_probability_derivative} show $\stabilityentropy$ and $\stabilitynorm$ averaged over the listening events of the users for $t>t_1$, \ie the time derivative of diversity and of the $\mathscr{l}^2$ norm of $\mathbf{p}^u(t)$ after the initial six-months period. Although $\stabilityentropy$ marginally increases for increasing values of $\levelofinterest$, we do not observe any significant difference among users of different $\levelofinterest$ in neither $\stabilityentropy$ nor $\stabilitynorm$. This is confirmed by the low correlation coefficients ($\rho=0.02$ for $\stabilityentropy(t)$ and $\rho=0.008$ for $\stabilitynorm(t)$) and indicates that, although users of higher $\levelofinterest$ explore more music \clusters{} and have more equally spread music tastes, the \textit{rate of change} of the relative amount of music listened per \cluster{} is roughly the same for all user groups. 
            
        \subsection{RQ 3: Time Exploration of Music \Clusters{}}
\begin{figure}
                \centering
                \begin{subfigure}[t]{0.48\textwidth}
                    \includegraphics[width=\columnwidth]{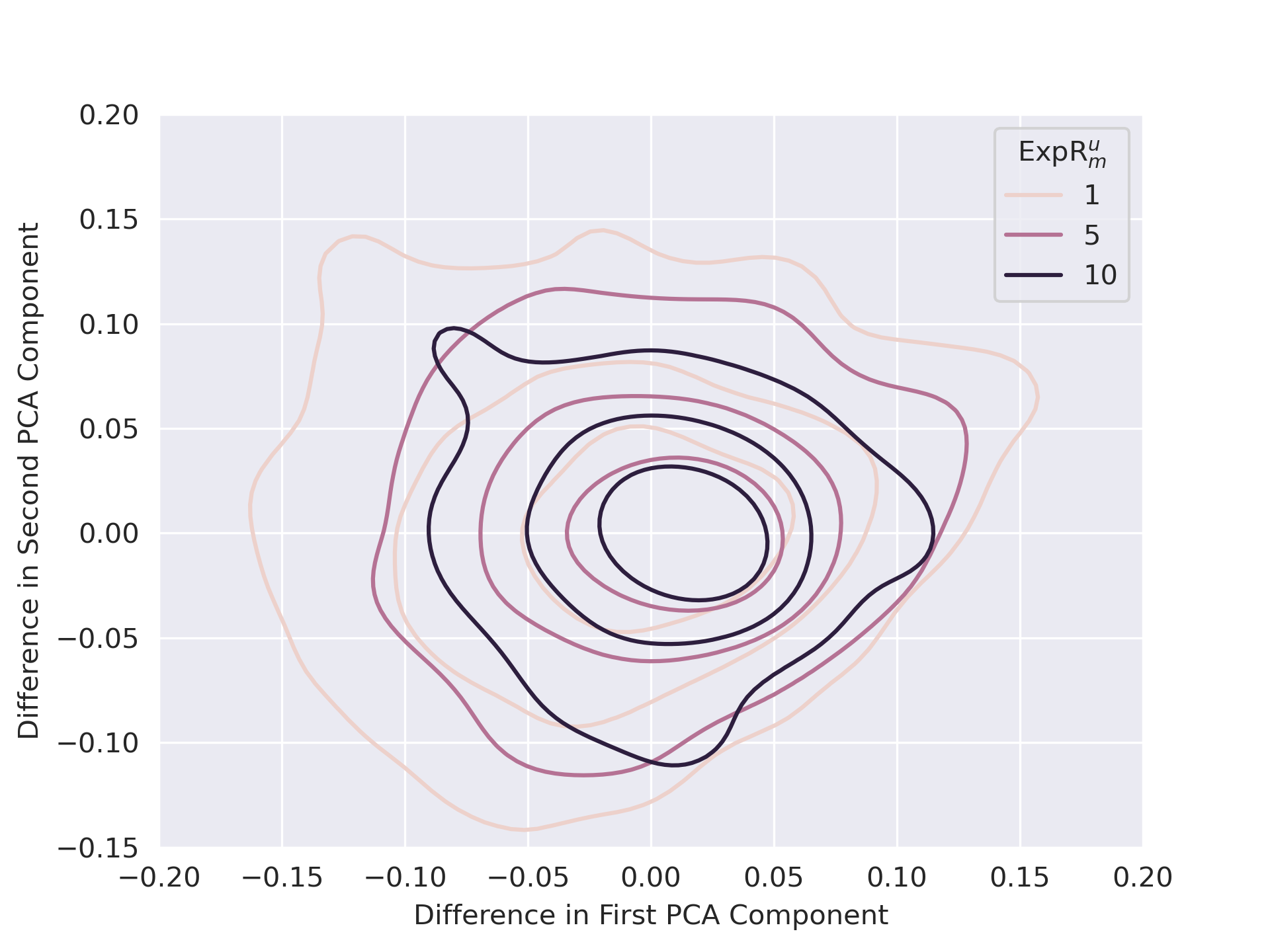}
                    \caption{Isolines for 0.2, 0.4, and 0.6\% of the distribution of distances from cluster centroids, for the first two principal components and for $\trackrank=1, 5, 10$. The distributions are approximated with KDE, as described in Section~\ref{sec:methodology_exploration_behavior}.}
                    \label{fig:rq3_pca_distance_representativeness_rank}
                \end{subfigure}
                \hfill
                \begin{subfigure}[t]{0.48\textwidth}
                    \includegraphics[width=\columnwidth]{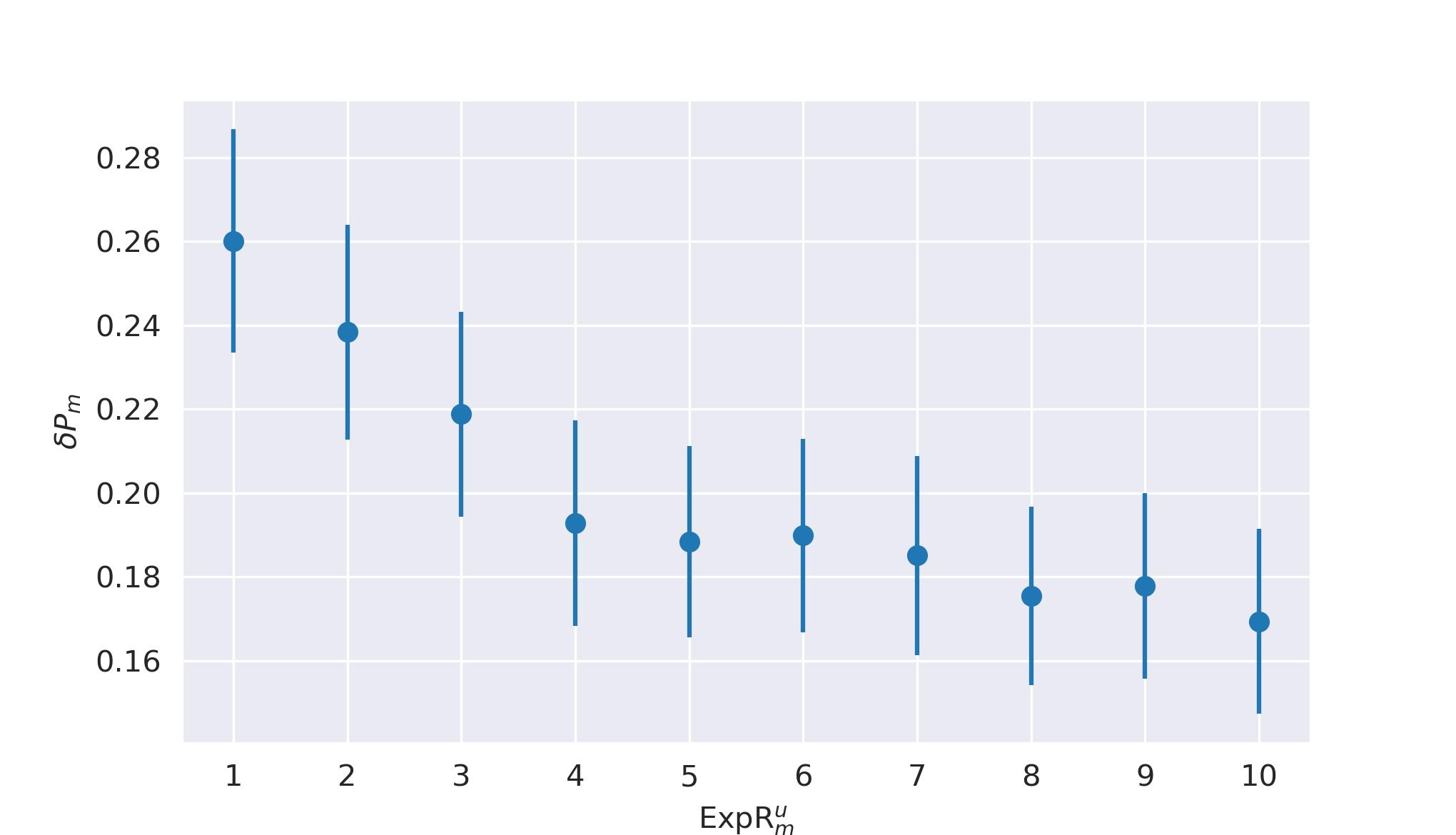}
                    \caption{Popularity $\relativepop$ over track time-rank $\trackrank$ in the \cluster{} exploration. Central dots and error bars represent mean and standard errors estimated with bootstrapping (see Section~\ref{sec:methodology_exploration_behavior}).}
                    \label{fig:rq3_popularity_rank}
                \end{subfigure}
                \caption{\Clusters{} exploration displayed as a function of the time-rank $\trackrank$ of tracks listened to in a newly discovered \cluster{}. These plots are computed over all users.}
                \label{fig:rq3_rank}
                \Description[]{\Cluster{} exploration displayed as a function of the time-rank $\trackrank$ of tracks listened to in a newly discovered genre. These plots are computed over all users.}
            \end{figure}

            Finally, we look at \textit{how} users explore newly discovered music \clusters{}. In particular, we look at \cluster{} representativeness and popularity of tracks newly listened after the \cluster{} discovery, and relative to tracks of the same \cluster{}. Figure~\ref{fig:rq3_pca_distance_representativeness_rank} shows, for the first two principal components, the KDE-approximated isolines of the distribution of distances from cluster centroid. As described in Section~\ref{sec:methodology_exploration_behavior}, these distributions are computed over all users, and for tracks of different rank $\trackrank$ in the exploration of a newly discovered \cluster{}, indicated by the different colors. Darker lines correspond to tracks explored later than lighter ones. To de-clutter the plot, we chose to display the isolines for $\trackrank=1,5,10$ only, \ie the first, the fifth, and the tenth listened tracks. We see that tracks that are listened later (isolines for $\trackrank=5$ in comparison to $\trackrank=1$, and of $\trackrank=10$ in comparison to $\trackrank=5$ and $\trackrank=1$) are on average closer to the cluster centroids. As revealed by the analysis on tracks in editorial playlists, cluster centroids seem to be more representative of the genres for the playlists investigated
            . Figure~\ref{fig:rq3_pca_distance_representativeness_rank} therefore indicates that after discovering a music \cluster{} (\ie after the first listen), users tend to listen to tracks that are more representative of the genre of the \cluster{}.  

            Figure~\ref{fig:rq3_popularity_rank} shows the popularity of tracks newly listened after the \cluster{} discovery and for different $\trackrank\in [1,2,\dots,10]$. The popularity $\relativepop$ is given as a relative number in relation to the average popularity of tracks of the same \cluster{}. We observe that $\relativepop > 0$ for all values of $\trackrank$. This indicates that in the initial phases of \cluster{} exploration, \ie for the first ten tracks, users listen to tracks that are more popular than the average 
            of the same \cluster{}. Furthermore, we observe that $\relativepop$ decreases as $\trackrank$ increases; more specifically, the plot indicates that users start exploring a \cluster{} from tracks that, on average, have a popularity $\sim26\%$ higher than the average of the \cluster{}, and later listen to tracks that are less popular than the starting one. The difference in $\relativepop$ between two consecutively listened tracks decreases as $\trackrank$ increases. This shows that users drift to less popular tracks with a faster rate at the beginning of the exploration of a \cluster{}.      
            The decrease in popularity and increase in \cluster{} representativeness (\ie decrease in $\relativedist$) during \cluster{} exploration are confirmed by the negative values of the correlation coefficients of $\relativepop$ and $\relativedist$ with $\trackrank$. ($\rho=-0.11$ for $\relativepop$ and $\rho=-0.15$ for $\relativedist$. In both cases $p<10^{-5}$.)

            The analyses summarized by Figure~\ref{fig:rq3_rank} are carried out over all users, irrespective of their self-reported interest in unfamiliar music. To investigate the correlation between self-reported interest in unfamiliar music $\levelofinterest$ and \cluster{} discovery patterns we perform two separate multivariate linear regressions, one of \cluster{} representativeness measured as $\relativedist$, and one of popularity $\relativepop$. Both regressions are performed on user interest in unfamiliar music $\levelofinterest$ and track time-rank $\trackrank$ in the \cluster{} exploration, considered as independent variables. The results of the regressions are reported in Table~\ref{tab:rq3_multivariate_popularity_distance}. First of all, the positive values of the intercepts for both $\relativepop$ ($0.314$, $p<0.001$) and $\relativedist$ ($0.108$, $p<0.001$) again indicate that users start exploring new \clusters{} from tracks that are more popular, and less representative (more distant from the centroid) than the average track in the \cluster{}. The negative slopes for $\trackrank$ ($-0.0150$ for popularity $\relativepop$, $-0.0359$ for distance from centroid $\relativedist$, in both cases $p<0.001$) again indicate that tracks that are listened later in the cluster exploration are less popular, but more representative of the music of the  explored \cluster{}. Finally, let us look at the estimated dependence on $\levelofinterest$, \ie on self-reported interest in unfamiliar music. The negative slope for $\relativepop$ (-$0.0105$, $p<0.001$) indicates that users of high levels of $\levelofinterest$ tend to explore \clusters{} through tracks that are less popular, compared to users with low $\levelofinterest$ levels. In contrast, the positive slope for $\relativedist$ ($0.0053$, $p<0.01$) indicates that tracks are less representative (more distant from the centroid) of the \clusters{} for users of higher $\levelofinterest$.   
            \begin{table}
                    \scalebox{0.86}{
                \begin{tabular}{r|r|r||r|r}
                     & \multicolumn{2}{c||}{$\relativepop$}&\multicolumn{2}{c}{$\relativedist$}\\
                     & Parameter & Std. Err.& Parameter & Std. Err. \\\hline 
                     Intercept & $0.3144^{***}$ & $0.0045$& $0.1076^{***}$ & $0.0068$ \\
                     $\trackrank$ Slope  & $-0.0150^{***}$ & $0.0007$& $-0.0359^{***}$ & $0.0011$ \\ 
                     $\levelofinterest$ Slope  & $-0.0105^{***}$ & $0.0013$& $0.0053^{**}$ & $0.0019$ \\
                \end{tabular}}
                \caption{Results of the multivariate linear regression of popularity $\relativepop$ and distance from cluster centroid $\relativedist$ on $\trackrank$ and $\levelofinterest$.  $^{***}p<0.001;^{**}p<0.01;^{*}p<0.05$.}
                \label{tab:rq3_multivariate_popularity_distance}
            \end{table}

\section{Conclusions, Limitations, and Future Work}
    \label{sec:conclusions}
        In summary, our work showed how music discovery patterns vary for users of different discovery needs. In particular, we showed that:
        \begin{description}[leftmargin=2mm,topsep=0pt]
            \item[RQ1] Users of higher declared interest in unfamiliar music explore more music 
            compared to those with lower declared interest.
            \item[RQ2] Users of higher declared interest in unfamiliar music exhibit more diverse music preferences than those with lower declared interest. 
            We did not observe any substantial difference in the stability of music preferences over time. 
            \item[RQ3] When users explore a new music \cluster{}, they start from tracks that are more popular and less representative of genres, compared to the average track of the same \cluster{}. While exploring, users gradually move to tracks that are less popular, but more representative of the \cluster{} genres. In addition, compared to those of lower declared interest, users of higher interest in unfamiliar music tend to explore music through tracks that are less popular and less representative of genres than the average track in the explored \cluster{}.
        \end{description}
    There are several limitations and potential extensions to our work. %
    First, the observation that the stability of users' music preferences over time does not vary over levels of interest in unfamiliar music seems to challenge previous findings on the relation between active engagement and stability of music preferences~\cite{liang2023promoting_music_exploration}. However, our analysis focused on interest in unfamiliar music \textit{specifically}, which is one of the seven aspects summarized by the overall active engagement level. Due to space limits, we did not extend the analysis to the other individual aspects. Future works could focus on the other aspects and identify the origin of the discrepancy. 
    The increasing representativeness in the exploration of a \cluster{} might be an indication that users are transitioning to a newly discovered \cluster{} from a previously known one, a hypothesis that could be tested in the future. The users sample consisted of users who replied to the survey and might therefore not be representative of the population of \deezer{} users (and of users of music streaming platforms more generally). In addition, our analysis uses a six-month window for the definition of \cluster{} discovery and long-term preferences. Although previous work on music streaming considered similar time-ranges~\cite{mehrotra2021spotify_familiarity,liang2023promoting_music_exploration}, this definition does not capture cases in which a user is familiar with a track but did not stream it for a long time, or in which a user familiarized with the track through other means, \eg listening to radio stations. The way we modeled exploration behavior considered the rank in the exploration phase, but did not consider aspects related to the time difference between consecutive discovers of a same \cluster{}, such as whether these tracks were explored in a same listening session.  We also did not distinguish between \textit{lean-back} and \textit{lean-in} streams, a distinction that might be particularly relevant to uncover exploration behavior.  Although the clustering is based on the track co-occurrence in user-generated playlists from \deezer{}, and although we showed that tracks that are in  \deezer{} editorial playlists are closer to cluster centroids, carrying out the analysis in terms of clusters instead of, \eg tag-based genres, renders some of the analyses more abstract. Finally, we carried out our analyses only in the music domain and on one dataset. This is due to the absence of other large-scale streaming dataset published together with users' replies to the Gold-MSI questionnaire.

    In addition to those mentioned above, extensions to our work could look at whether \cluster{} exploration and repeated consumption of tracks from a same \cluster{} show the same trends of satiation as for repeated consumption of individual music tracks, extensively analyzed in past works~\cite{reiterhaas2021relistening,sguerra2022discovery_dynamics}. Finally, one could leverage the insight gained in this work to develop RSs that guide users in exploring music by adjusting the levels of popularity and of genre representativeness of recommended tracks. By means of a user study one could further test if adjusting these aspects to the users' self-reported interest in unfamiliar music leads to a higher satisfaction with recommendations for music exploration.

    Our work opens up possibilities to infer users' interest in unfamiliar music from streaming data, as well as to develop RSs that guide users in exploring music in a more natural way by considering popularity and genre representativeness of recommendations.

\begin{acks}
    This research was funded in whole or in part by the Austrian Science Fund (FWF) \url{https://doi.org/10.55776/P33526}, \url{https://doi.org/10.55776/DFH23}, \url{https://doi.org/10.55776/COE12}, \url{https://doi.org/10.55776/P36413}, and by the French National Agency of Research Grant (ANR) “RECORDS” (ANR-2019-CE38-0013). 

\end{acks}
\balance
\bibliographystyle{ACM-Reference-Format}
\input{main.bbl}

\end{document}